\documentclass[pre,aps,pacs,twocolumn]{revtex4}
\usepackage{epsfig}
\usepackage{bm}
\usepackage{color}

\newcommand{\C}[1]{{\mathcal{#1}}}
\usepackage[latin1]{inputenc}
\newcommand{\beq}{\begin{equation}}
\newcommand{\eeq}{\end{equation}}
\newcommand{\bea}{\begin{eqnarray}}
\newcommand{\eea}{\end{eqnarray}}

\begin{document}
\title{Statistical Mechanics of the Glass Transition}
\author{H. G. E. Hentschel$^1$, Valery Ilyin$^2$,  Nataliya Makedonska$^2$, Itamar Procaccia$^2$ and Nurith Schupper$^2$}
\affiliation{$^1$ Department of Physics, Emory University, Atlanta,  Georgia 30322\\
$^2$The Department of Chemical Physics, The Weizmann
Institute of Science, Rehovot 76100, Israel, }
\begin{abstract}
 The statistical mechanics of simple glass forming systems in 2 dimensions is worked out.
 The glass disorder is encoded via a Voronoi tessellation, and the statistical mechanics is performed
 directly in this encoding.  The theory provides, without free parameters,  an explanation of the glass transition phenomenology, including the identification of two different temperatures, $T_g$ and $T_k$, the first associated with jamming and the second associated with the appearance of a quasi-crystal at very low temperatures.  
  \end{abstract}
\maketitle

{\bf Introduction:} The term ``glass transition" refers to the enormous slowing down in the dynamics of some liquids when their temperature is lowered. Despite decades of research, a clear explanation of this phenomenon, common to materials as diverse as molecular glasses, metallic glasses, colloids etc., is still lacking \cite{01Don, 01DS}. The difficulty is that the disordered molecular arrangement in a glass appears indistinguishable from that of the corresponding liquid, without any sign of  a static correlation length that increases appreciably at the glass transition \cite{05BBBCELLP}. In this Letter we offer a theory of the glass transition in simple glass forming systems, based on an encoding of the disorder that is able to flush out the pertinent features of the transition. In particular the disappearance of liquid-like regions, the huge increase of a typical scale, and eventually the existence at low temperatures of a 
quasi-crystalline state that blocks true crystallization \cite{01CDZIS,05Tan}, are all understood, in agreement with recent numerical simulations. 
\begin{figure}
\centering
\epsfig{width=.35\textwidth,file=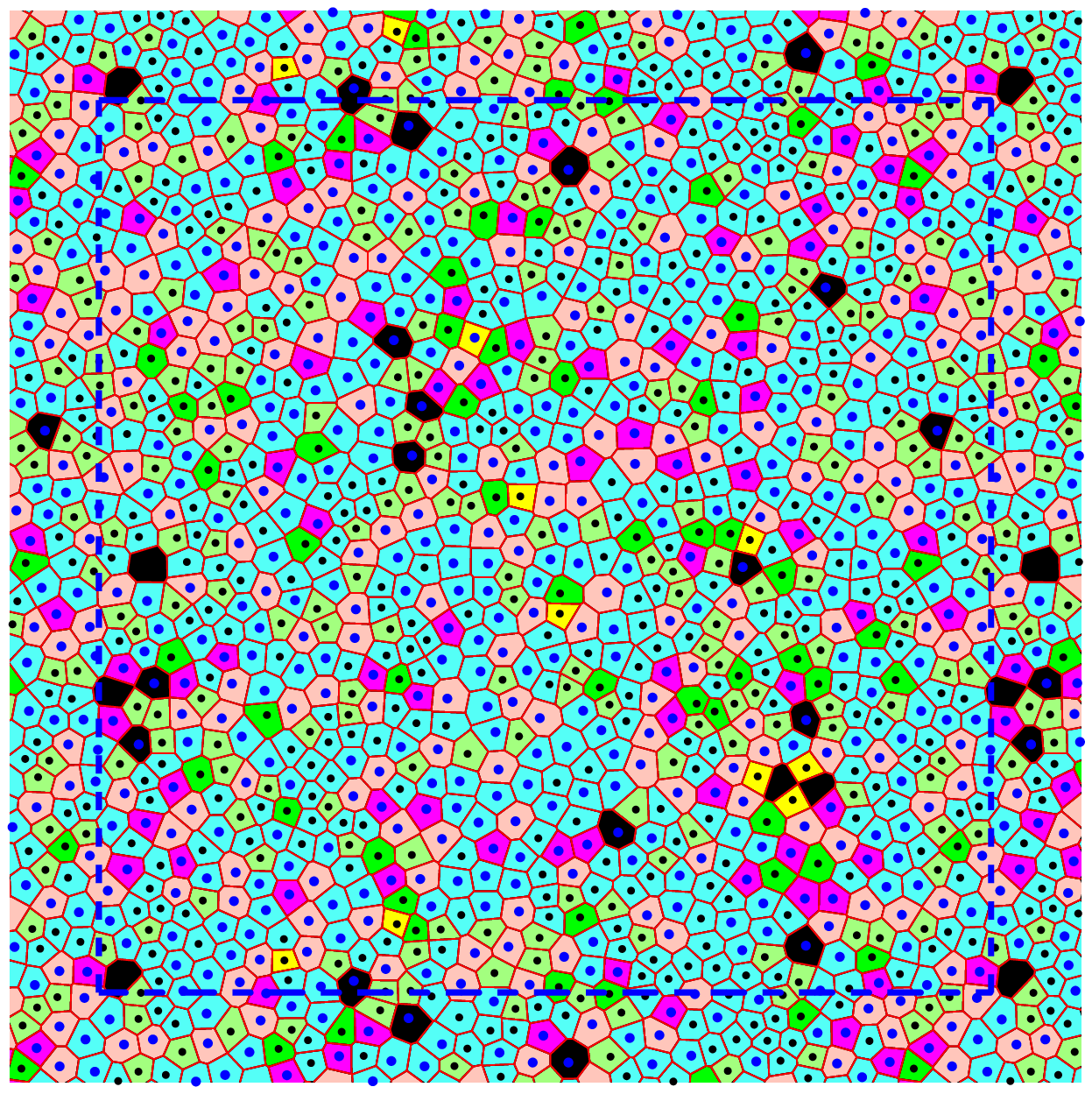}
\epsfig{width=.35\textwidth,file=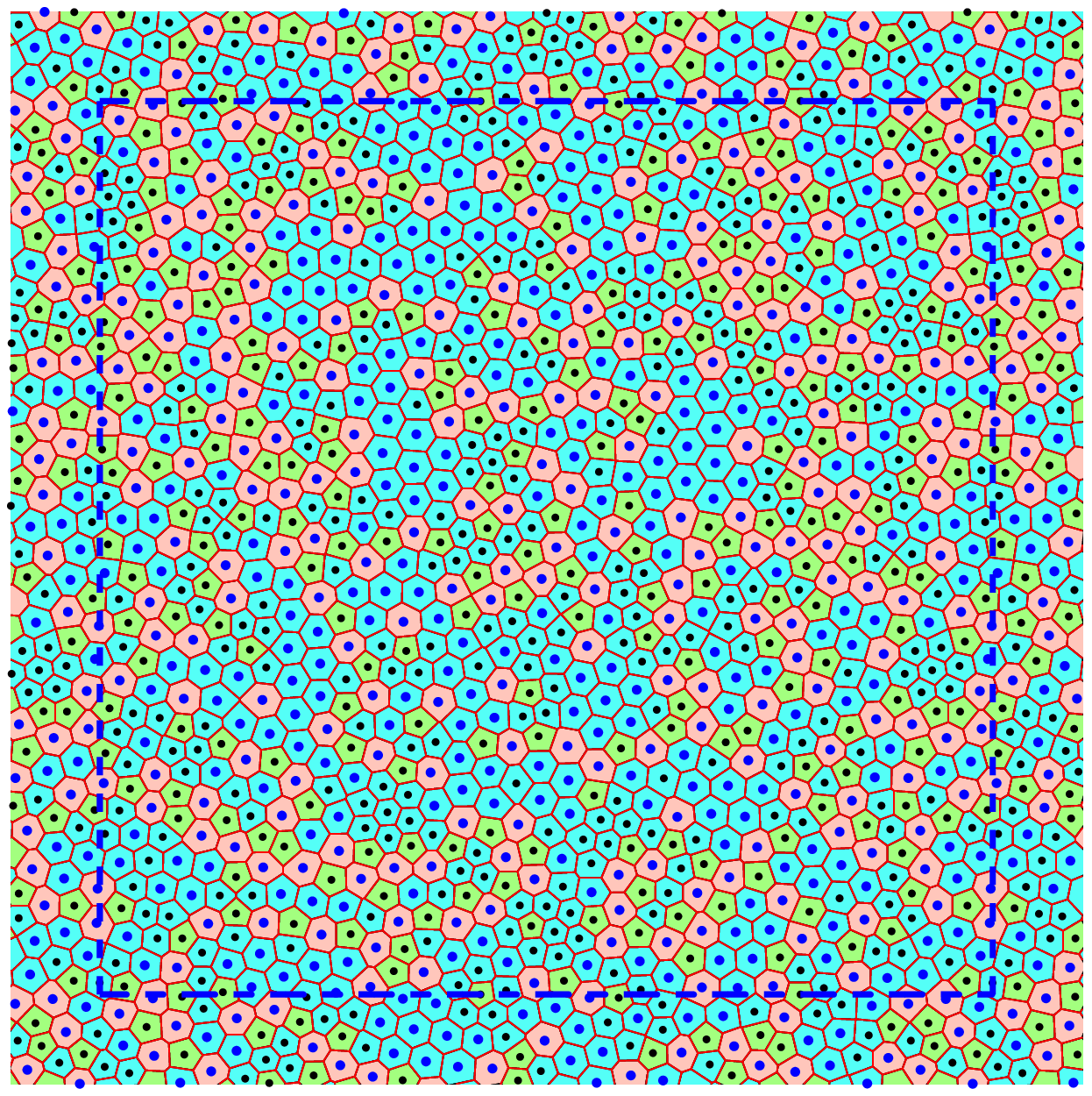}
\caption{Upper panel: a Voronoi polygon construction at $T=3$,  with the seven-color code used in this paper. Small particles in pentagons (heptagons) are light green (dark green) and large particles in pentagons (heptagons) are violet  (pink). Hexagons are in blue, octagons in black and squares in yellow. Lower panel: a typical
Voronoi construction in the glass phase at $T=0.1$. Note the total disappearance of liquid-like
defects.}
\label{Voronoi}
\end{figure}

{\bf The system}: the glass former explored here is the well studied system of a two-dimensional binary mixture of discs interacting via a soft $1/r^{12}$ repulsion with a ``diameter" ratio of 1.4.  We refer
the reader to the extensive work done on this system \cite{99PH}, showing that it is a {\em bona fide} glass-forming liquid meeting all the criteria of a glass transition. The system consists of an
equimolar mixture of large particles with `diameter' $\sigma_2=1.4$ and small particles of `diameter' $\sigma_1=1$, respectively, but with the same mass $m$. The three pairwise additive interactions are given by the purely repulsive soft-core potentials
\begin{equation}
u_{ab} =\epsilon \left(\frac{\sigma_{ab}}{r}\right)^{12} \ , \quad a,b=1,2 \ , \label{potential}
\end{equation}
where $\sigma_{aa}=\sigma_a$ and $\sigma_{ab}= (\sigma_a+\sigma_b)/2$. The cutoff radii of 
the interaction are set at $4.5\sigma_{ab}$. The units of mass, length, time and temperature are $m$, $\sigma_1$, $\tau=\sigma_1\sqrt{m/\epsilon}$ and $T=\epsilon/k_B$, respectively, with $k_B$ being
Boltzmann's constant. A total of $N=1024$ particles were enclosed in a
square box (of area $L^2$) with periodic boundary conditions.  Ref. \cite{99PH} found that for temperature $T>0.5$ the system is liquid and  for lower temperatures dynamical relaxation slows down. A precise glass transition had not been identified in \cite{99PH}.

{\bf The glass transition:} To overcome the lack of signatures of the glass transition in the particle
positions on the molecular level, we encode the state of the system using the 
Voronoi tesselation \cite{89DAY}, where a polygon associated with any particle contains all points closest to that particle than to any other particle. The edges of such a polygon are the perpendicular bisectors of the vectors joining the central particle. Such an encoding has been used extensively before,  \cite{89DAY,99PH}, where it was also noted that the geometric Euler constraint implies that average coordination number is 6 at all temperatures, and local coordination numbers other than 6 were referred to as ``defects". Our encoding is richer; In our work \cite{06ABIMPS} it was discovered that a significant insight to the glass transition is gained by distinguishing between ``liquid-like" defects and ``glass-like" defects. We observed \cite{06ABIMPS} that only in the liquid phase there exist {\bf small particles} enclosed in heptagons (or even octagons), and  {\bf large particles} enclosed in pentagons (or even squares) (cf. Fig. \ref{Voronoi} upper panel). In the glass phase we observe only defects of the opposite type, i.e. small particles in pentagons and large particles in heptagons, see Fig. \ref{Voronoi}. Accordingly we proposed
that the concentration of these particular defects is a good indicator of the glass transition.
 {\bf The concentration $c_\ell$ of these liquid-like defects becomes so small in the glass phase that we cannot distinguish it from zero} (cf. Fig. \ref{Voronoi} lower panel), unless the glass is put under mechanical strain, cf. Ref. \cite{06ABIMPS}. 
\begin{figure}
\centering
\epsfig{width=.35\textwidth,file=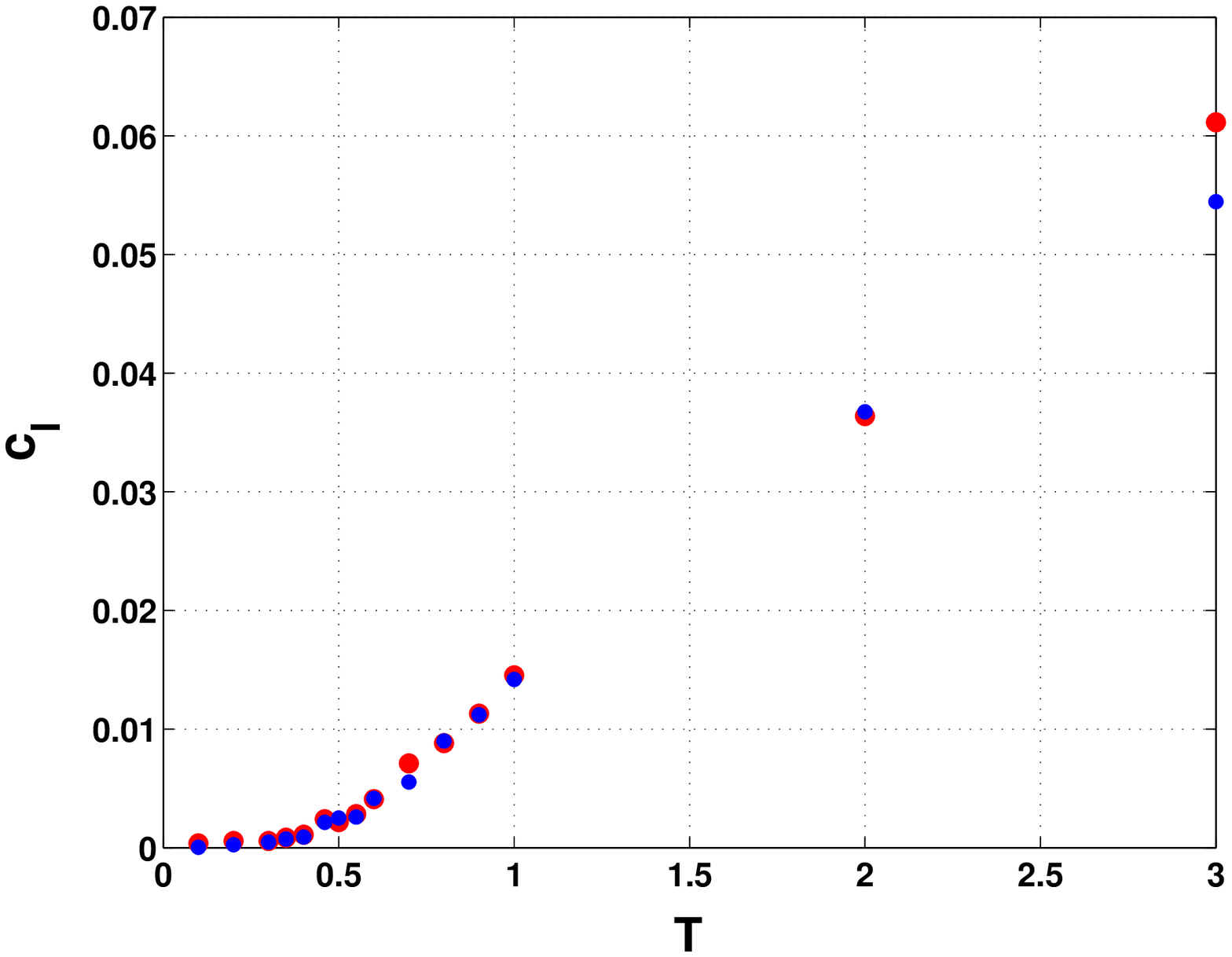}
\epsfig{width=.35\textwidth,file=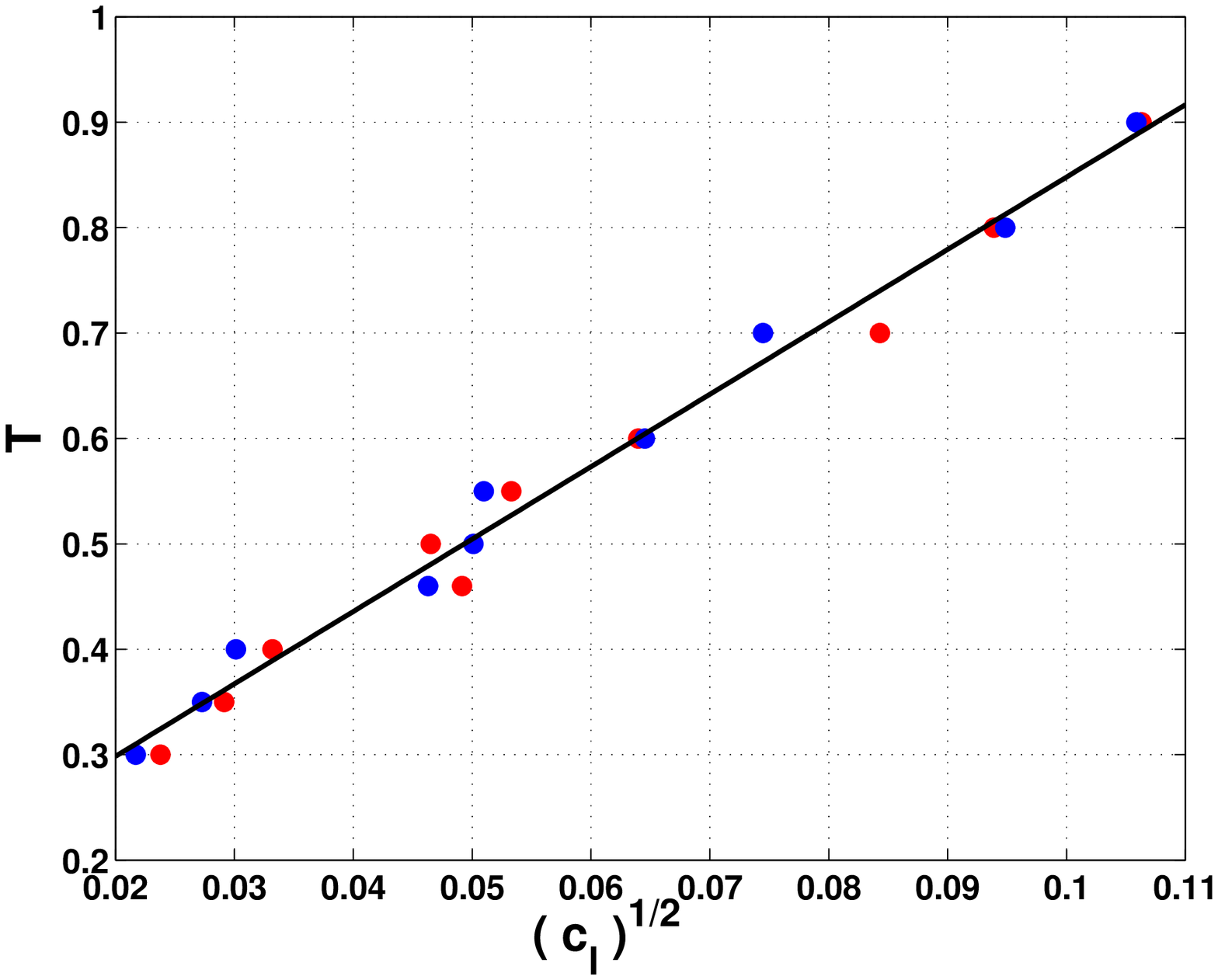}
\caption{Concentration of defects under slow cooling. Upper panel: the concentration of the
liquid-like defects: large particles in
pentagons (red dots) and small particles in heptagons (blue dots). 
Lower panel: the fit of the concentration of liquid-like defects as a function of temperature according
to Eq. (\ref{PLfit}).}
\label{concentrations}
\end{figure}
Associated with this concentration we can define a typical scale, $\xi$:
 \begin{equation}
 \xi \equiv \case{1}{\sqrt{c_\ell}} \ . \label{defxi}
 \end{equation}
Associated with the strong decrease in $c_\ell$ we observe a huge increase in the typical scale $\xi$,
in agreement with the tremendous slowing down of the dynamics. 

In Fig.~ \ref{concentrations} upper panel we show $c_\ell$ as a function of the temperature for a protocol of slow cooling. For temperatures larger than 0.8 the concentration follows closely an exponential fit,
\begin{equation}
c_\ell = A \exp{-(\Delta U/T)} \ , \quad A\approx 0.094, ~\Delta U\approx 1.90 \ .  \label{expfit}
\end{equation}
For temperatures in the range $0.3<T<0.8$ we find an excellent fit to
\begin{equation}
c_\ell = B (T-T_g)^2\  , \quad  B\approx 0.02,~T_g=  0.16 \pm 0.02\ .\label{PLfit}
\end{equation}
The quality of this fit is demonstrated in  Fig. \ref{concentrations} lower panel. 
The fit (\ref{PLfit}) appears to identify a sharp glass transition $T_g=0.16\pm 0.02$; note, however, that the fit is made in the range $0.3<T<0.8$ that does not include $T_g$. In fact there is no theoretical reason to expect that $c_\ell$ truly
vanishes at $T_g$, but it becomes so small that we indeed do not see a single liquid-like defect
in our finite-box simulations. Similarly, in our finite size box we cannot distinguish between a
diverging length-scale and a scale much larger than $L$.  Within the range $0.3<T<0.8$ we fit
an apparently divergent length
\begin{equation}
\xi \sim (T-T_g)^{-\nu} \ , \quad \nu=1 \ . \label{xi}
\end{equation}
\begin{figure}
\centering
\epsfig{width=.35\textwidth,file=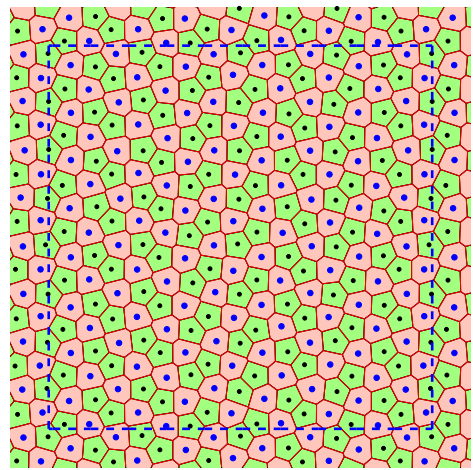}
\caption{The `ideal' glass state.}
\label{quasicrystal}
\end{figure}
Finally, it was discovered in \cite{06ABIMPS} that at sufficiently low temperatures, there exists
a quasi-crystalline phase, see. Fig. \ref{quasicrystal}.  This phase arises while heating, starting
from a square lattice with interchanging small and large particles. The quasi-crystalline phase appears stable in our slow heating simulations as long as the temperature is sufficiently small. We propose to identify the quasi-crystalline phase as the ideal glass state, and the temperature where it melts as the Kauzmann temperature $T_k$ \cite{48Kau}. Upon heating above $T_k$ the quasi-crystal melts in favor of a generic glassy state in which the quasi-crystalline phase is
invaded by patches of hexagons, see Fig.~\ref{Voronoi} lower panel. To understand these findings
we need to understand why the liquid-like defects disappear first upon cooling, and what is the
thermodynamic reason for the non-existence of hexagons at sufficiently low temperatures.

\begin{figure}
\centering
\epsfig{width=.40\textwidth,file=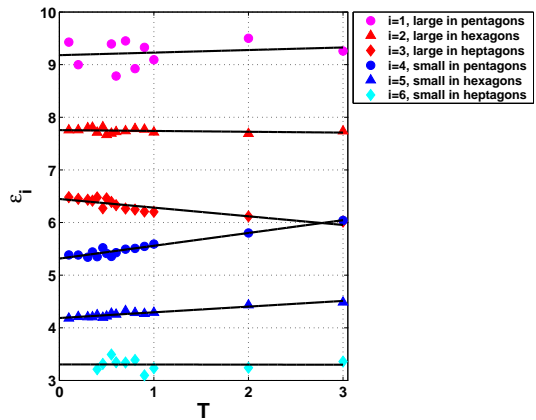}
\caption{The average energies of the voronoi cells as a function of the temperature as measured in the simulations.}
\label{energies}
\end{figure}

{\bf Statistical Mechanics:} here we present the statistical mechanics of this system, rationalizing
and explaining the features of the transition presented above. The main point to stress is that 
even in the glass phase, when particles are practically jammed and precluded from large
excursions, the Voronoi tesselation is still highly mobile. Minute changes in particle positions
can lead, via T1 foam-like processes \cite{99WH} to transitions between cell-types. Therefore, if one
can convince oneself that the cell types, including the presence of a small or large particle
inside them,  are proper `species', one can construct statistical
mechanics right on the space of these cells. We define the energy of each cell type as the average (over all cells of the same type) of the potential energy 
 $ \epsilon_i=\langle \sum_{k=1}^{E_i}\epsilon \left(\frac{\sigma_{ik}}{r_{ik}}\right)^{12}\rangle $ where $E_i$ is the
 number of edges associated with that cell type, and
$r_{ik}$ being the distance to the particle in the adjacent Voronoi cell. In 
Fig. \ref{energies} we present the values of these energies measured 
as a function of the temperature, following a protocol of slow cooling. There are 10 different cell types in this system, (large paricle or small particle in squares, pentagons, hexagons, heptagons and octagons), but octagons and squares are not shown since they disappear much before the glass transition. The lesson drawn from this graph is that {\bf the different cell types have clearly
split energies throughout the interesting temperature range, and that these energies are only
weakly dependent on the temperature}. Within the temperature range of interest we can focus on the six types of cells; denote by  $\{N_i\}_{i=1}^6$ the number of cells of each type, with number of edges $E_i$, ordering them by the energy $\epsilon_i$ from $i=1$ being the highest one (large particle in a pentagon) to $i=6$ being the small particle in an heptagon. Additional important properties of the cell types are their areas $\Omega_i$ and their shapes; the first affects the enthalpy term  and both affect the entropy when we count the number of possible tilings of the plane.
With this in mind we can construct the statistical mechanics of this system by considering
the free energy $G=U+pV-TS$.  The value of $U$ is then simply 
$\sum_{i=1}^6 N_i \epsilon_i$. The $pV$ term is simply $p\sum_{i=1}^6 N_i \Omega_i$. 
 Lastly, we need to estimate the entropy term. In principle this should be computed from the number
 of possible complete tilings of the area by $N_i$ cells of each type with its given area
 and shape, subject to the Euler constraint $\sum_{i=1}^6 N_i E_i =6N$. This is a formidable problem. A useful estimate can be made by considering the area
 only, and filling space starting with the largest objects, then the next largest etc. until the smallest
 is fit in. Denoting the possible number of boxes to fit the largest cells by ${\cal N}_1\equiv V/\Omega_1$, then the number of boxes available for the second largest cell by $\C N_2\equiv
 (V-N_1\Omega_1)/\Omega_2$ etc., the number of possible configurations $W$ is
 \begin{equation}
 W = \Pi_{k=1}^6 \frac{\C N_k!}{N_k!(\C N_k-N_k)! }\ . \label{W}
 \end{equation}
Denoting $x_i\equiv N_i/\C N_i$ we compute directly $x_i=c_i\Omega_i/\sum_{j=i}^6 c_j\Omega_j$
where $c_i$ is the number concentration of each defect. We can now compute $S=k_{\rm B}\ln W$ and write $G$ together with a Lagrange multiplier for the Euler constraint,
\begin{eqnarray}
&&G = \sum_{i=1}^6 N_i \epsilon_i + p\sum_{i=1}^6 N_i \Omega_i +\lambda \sum_{i=1}^6 N_i E_i  \nonumber\\
&&+T\sum_{k=1}^6 \C N_k[x_k \ln x_k+(1-x_k)\ln (1-x_k)]  \ . \label{G}
\end{eqnarray}
The chemical potential $\mu_i \equiv \partial G/\partial N_i$ is then
\begin{equation}
\mu_i =\epsilon_i +p\Omega_i +T[\ln x_i +\sum _{k=1}^{i-1} \case{\Omega_i}{\Omega_k}\ln(1-x_k)]+\lambda E_i \ . \label{general}
\end{equation}
\begin{figure}
\centering
\epsfig{width=.35\textwidth,file=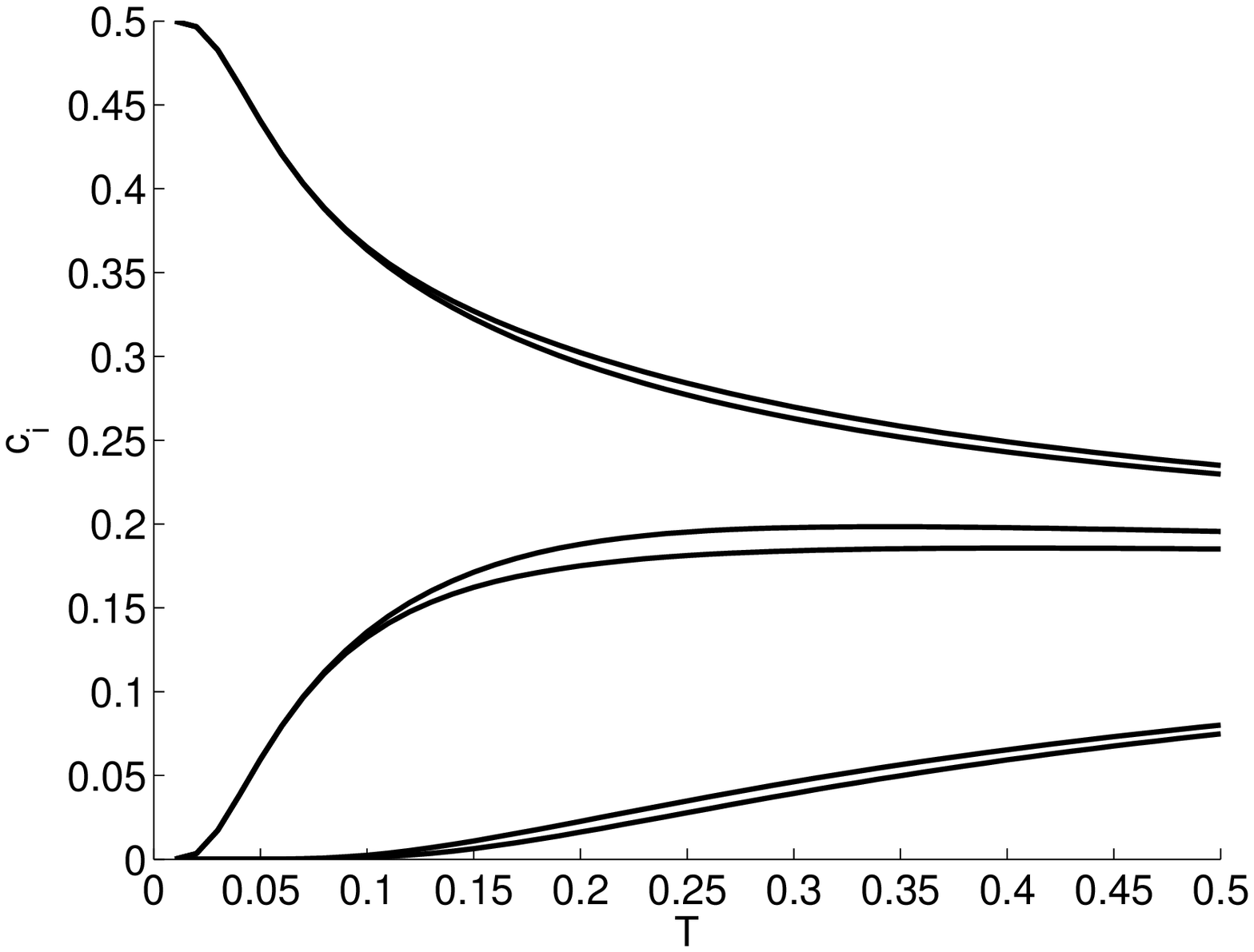}
\epsfig{width=.35\textwidth,file=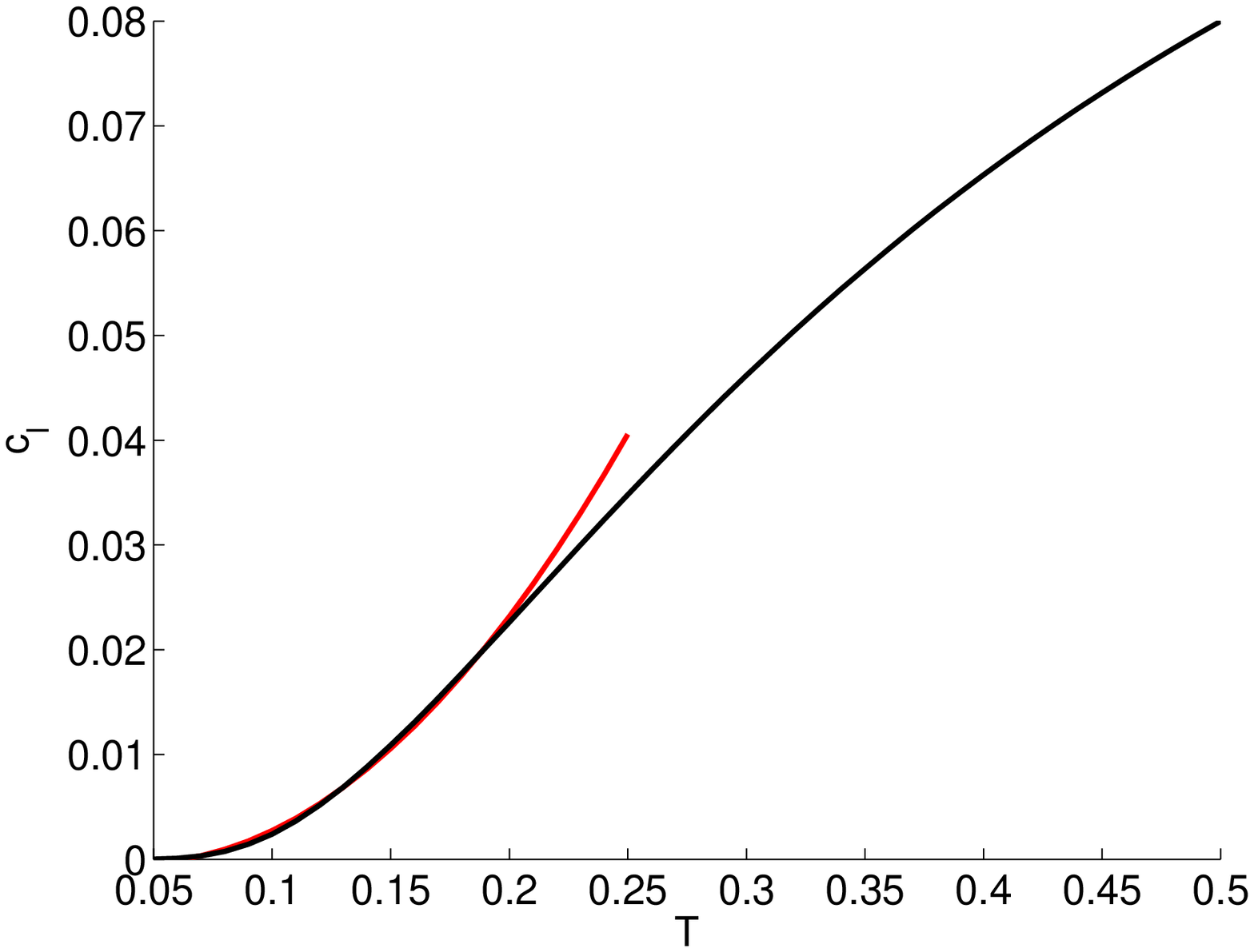}
\caption{The concentrations of Votonoi cells as predicted by the statistical mechanics. Note the close-to-degeneracy of pairs of cells: the upper pair are the glass-like defects, the middle pair are the two
hexagons and the lower pair the liquid-like defects. Observe the existence of two temperatures, interpreted as $T_g$ and $T_k$.
Lower panel: the concentration of the liquid-like defect (small in heptagon) as a function of temperature,
together with the quadratic fit of the type Eq. (\ref{PLfit}).}
\label{statmech}
\end{figure}

We now recognize that in equilibrium there exist only two independent values of $\mu_i$, one
for the small particles $\mu_S$  and one for the large particles $\mu_L$, and we have 9 unknowns --
six values of $c_i$, 2 values of $\mu$ and one Lagrange multiplier $\lambda$. This is precisely
balanced by the 6 equations (\ref{general}), the Euler constraint, and the two constraints $\sum_{i=1}^3 c_i=\sum_{k=4}^6 c_k=1/2$. These equations can be solved numerically
using the precise values of $\Omega_i(T)$ and $\epsilon_i(T)$ as measured in the simulation.
The approximate calculation of the entropy however does not warrant such a detailed calculation.
In reality, calculating the average areas of the cell types in the numerical simulations, we discover
that to an excellent approximation these fall in two classes, smaller cells of area $\Omega_S$ when
small particles are in them, and larger cells of area $\Omega_L$ when large particles are enclosed.
These areas again are only weakly dependent on the temperature. Then the whole system of
equations simplifies to two analytically tractable sets of equations
\begin{eqnarray}
\tilde \mu_L &=& \epsilon_i + T\ln c_i +\lambda E_i \ , \quad \{i=1,2,3\} \ , \nonumber\\
\tilde \mu_S &=& \epsilon_i + T\ln c_i +\lambda E_i \ , \quad \{i=4,5,6\} \ , \label{simple}
\end{eqnarray} 
together with the above mentioned three constraints. In $\tilde \mu$ we have absorbed terms that
added to $\mu$ in this special case. In Fig. \ref{statmech} we show the solutions of these equations
when we use values of $\epsilon_i$ taken from the Fig. \ref{energies} at $T=0$ \cite{foot}. The results
are in excellent  agreement with the simulations. There is a clear degeneracy in the temperature dependence of the two types of liquid like defects, the two hexagons and the pair of glass-like defects. At low temperatures we only have the glass-like defects who cannot crystalize, but can form the quasi-crystalline phase. Upon warming up, at $T_k$, a sizable
number of hexagons appears to form the generic
glassy state. Yet later, at $T_g$, a sizable number 
of liquid-like defects brings the system to a liquid state. The curves associated with the temperature dependence of the concentration of the latter defects can be very well fit
by a quadratic fit like Eq. (\ref{PLfit}), with a temperature $T_g\approx 0.07$. In fact, the solution of Eqs. (\ref{simple}) shows that
the concentrations do not quite vanish but become exponentially small, as stated above and in \cite{06ABIMPS}. Note the numerical value of $T_g$ differs from the simulation, as is expected
from a theory that does not take careful account of the correlations between different cells. The actual values
of $T_g$ and $T_k$ can be understood from this model. Denote by $c_\ell$, $c_H$ and $c_G$ the concentrations of the liquid-like, hexagons and glass-like defects, and by $\epsilon_\ell=\epsilon_1+\epsilon_6\approx 12.48$ as the energy associated with the liquid-like defects, by $\epsilon_H=\epsilon_2+\epsilon_5\approx 11.94$ the energy of the
hexagons, and $\epsilon_G=\epsilon_3+\epsilon_4\approx 11.76$ the energy of the glass-like defects. The
theory predicts that ratios $c_\ell/c_H$ and $c_H/c_G$ are of the order of $\exp[-(\epsilon_\ell -\epsilon_H)/T ]$ and 
$\exp[-(\epsilon_H -\epsilon_G)/T ]$ respectively. As an estimate of $T_g$ and $T_k$ take these ratios to be, say,
of the order of $ 1\%\sim \exp(-5)$ and observe that such ratios are obtained for $T=T_g\approx 0.11$ and $T=T_K\approx 0.04$. It is important to notice that $\epsilon_H -\epsilon_G$ could be positive rather than negative, and then the system would crystallize on a hexagonal lattice!

 {\bf Conclusions:} It is quite remarkable that one can understand the properties
 of a system of strongly interacting particles in an essentially jammed state using the statistical mechanics
 of a mobile set of essentially non-interacting quasi-particles. Without any free parameter we have offered a semi-quantitative understanding of the features of the glass transition in this system, as well as the existence of two distinct temperatures $T_g$ and $T_k$.  

\acknowledgments
This work had been supported
in part by the Israel Science Foundation and by the German-Israeli Foundation.


\begin{thebibliography}{99}
\bibitem{foot}
Endowing the individual energies with the measured temperature dependence did not result
in an important change in the statistical mechanics.
\end{thebibliography}
\end{document}